\documentclass[aps,prl,times,twocolumn,amsmath,amssymb,superscriptaddress,floatfix,showpacs]{revtex4}
\usepackage{color}

\usepackage[dvips]{graphicx}
\usepackage{subfigure}
\usepackage{bbold}
\usepackage{verbatim}
\usepackage{float}
\usepackage{dsfont}
\begin{document}
\title{Supersolid phases in a realistic three-dimensional spin model}
\author{Luis Seabra}
\affiliation{H.\ H.\ Wills Physics Laboratory, University of Bristol,  Tyndall Av, BS8--1TL, UK.}
\author{Nic Shannon}
\affiliation{H.\ H.\ Wills Physics Laboratory, University of Bristol,  Tyndall Av, BS8--1TL, UK.}

\date{\today}
\begin{abstract}   
Supersolid phases, in which a superfluid component coexists with conventional crystalline long range order, 
have recently attracted a great deal of attention in the context of both solid helium and quantum spin 
systems.  Motivated by recent experiments on $2H$-AgNiO$_2$, we study the magnetic phase diagram of 
a realistic three-dimensional spin model with single-ion anisotropy and competing interactions on a layered 
triangular lattice, using classical Monte Carlo simulation techniques, complemented by spin-wave calculations.
For parameters relevant to experiment,  we find a cascade of different phases as a function of magnetic field, 
including three phases which are supersolids in the sense of Liu and Fisher.  One of these phases is continuously 
connected with the collinear ground state of AgNiO$_2$, and is accessible at relatively low values of magnetic field.   
The nature of this low-field transition, and the possibility of observing this new supersolid phase 
in AgNiO$_2$, are discussed. 

\end{abstract}

\maketitle

%%%%%%%%%%%%%%%%%%%%%%%%%%%%%%%%%%%%%%%%%%%%%%%%%%%%%%

Solids and liquids are very different.   Placed under stress, a liquid will flow, while a solid resists deformation.    
The idea of a supersolid, a state which combines the properties of a solid with those of a perfect, non-dissipative superfluid, 
seems therefore to fly in the face of common sense.   None the less, the proposal that a supersolid might occur through the 
Bose-Einstein condensation of vacancies in a quantum crystal\cite{andreev69}, was propelled to the centre of debate  
by recent experiments on $^4$He\cite{chan04}.

A radically different approach to supersolids was initiated by Liu and Fisher\cite{liu73}, who realised that 
quantum magnets could support states which break the translational symmetry of the lattice (and are therefore solids)
while {\it simultaneously} breaking spin-rotational symmetry within a plane, a form of order analogous to a superfluid.
It is now well established that models of two-dimensional 
frustrated magnets with easy-axis anisotropy can support such supersolid states\cite{melko05}.   Moreover, since the states 
of a spin-1/2 quantum magnet are in one-to-one correspondence with hard-core bosons, these supersolids might also 
be realised using cold atoms on optical lattices.  
Nonetheless, candidates for supersolid states among real, three-dimensional magnets remain scarce.   
An interesting system in this context is the triangular easy-axis magnet, $2H$-AgNiO$_2$\cite{wawrzynska07}.

\begin{figure}[ht]
\centering
\includegraphics[totalheight=0.15\textheight]{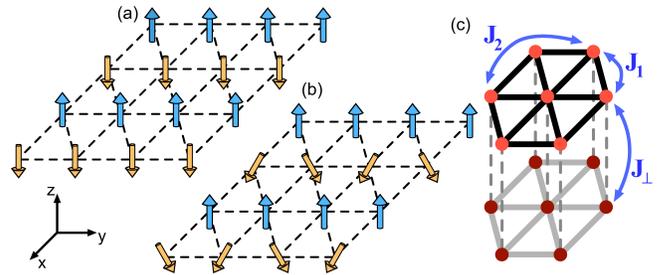}
\caption{\footnotesize{(Color online) a)  Low-field collinear ``stripe'' ground state, with spins aligned along the 
magnetic easy axis (z-axis).    b) Related supersolid phase for magnetic field parallel to the easy axis~: ``down" spins cant into plane
perpendicular to field, while ``up'' spin remain aligned with the field.   
c) First-neighbour $J_1$, second-neighbour $J_2$, and interlayer interactions $J_\perp$ for a stacked a triangular lattice.}}
\label{cartoon}
\end{figure}

AgNiO$_2$ is a very unusual material, built of stacked, two-dimensional nickel-oxygen planes, held together by silver ions. It combines metallicity and magnetism, with the magnetic ions in each plane forming a perfect triangular lattice, nested within a honeycomb network of conducting sites\cite{wawrzynska07}.   
In the absence of magnetic field AgNiO$_2$ supports a stripe-like collinear antiferromagnetic ground state, illustrated in Fig.~\ref{cartoon}(a).   
Recently, AgNiO$_2$ has been shown to undergo a complicated set of phase transitions as a function of magnetic 
field \cite{coldea10}.  Of particular interest is the transition out of the collinear ground state at low temperatures.   

In applied magnetic field, collinear antiferromagnets with easy-axis anisotropy typically undergo a first-order ``spin-flop'' 
transition into a canted state, at a critical field which is broadly independent of temperature.   However the low-field transition in 
AgNiO$_2$ is accompanied by a relatively broad feature in specific heat, does not exhibit marked hysteresis, and occurs at 
progressively higher fields as temperature increases. 
None of these features resemble a typical spin-flop transition, and together they raise the question of whether a novel type of 
magnetic order is realised in AgNiO$_2$ under field.

In this Letter we explore the different phases that occur as a function of magnetic field in a simple effective spin model already 
shown to provide excellent fits to inelastic neutron scattering spectra for AgNiO$_2$\cite{wheeler09}.   We show that the 
collinear ground state of this model does not undergo a conventional spin-flop, but rather a Bose-Einstein condensation of magnetic excitations 
which converts it into it a state that is a supersolid in the sense of Liu and Fisher.    We also identify two magnetization plateaux, 
and two further supersolid phases at high field.

The model we consider is the Heisenberg model on a layered triangular lattice, with competing antiferromagnetic
first- and second-neighbour interactions $J_1$ and $J_2$, single-ion anisotropy $D$ and inter-layer coupling $J_\perp$
\begin{eqnarray} 
\label{eq:H}
\mathcal{H} =& J_1\sum_{\langle ij \rangle_1} {{\bf S}_i} \cdotp {\bf S}_j + J_2\sum_{\langle ij \rangle_2} {\bf S}_i \cdotp {\bf S}_j
+ J_\perp\sum_{\langle ij \rangle_\perp} {\bf S}_i \cdotp {\bf S}_j  \nonumber\\
& -D \sum _i ({{ \bf S}_i^z})^2   - h\sum_{i}{\bf S}^z_i. 
\end{eqnarray} 
(cf.~Fig.~\ref{cartoon}(c)).  
For concreteness, we set $J_1=1$, $J_2=0.15$, $J_\perp=-0.15$ and $D=0.5$, measuring 
magnetic field $h$ and temperature $T$ in units of $J_1$.    These are ratios of parameters comparable to those used to 
fit inelastic neutron scattering spectra for AgNiO$_2$\cite{wheeler09}.  
Like AgNiO$_2$, in the absence of magnetic field, this model exhibits a collinear ``stripe-like'' magnetic ground state, illustrated
in Fig.~\ref{cartoon}(a).     The stripes have three possible orientations, and so break a $\mathds{Z}_3$ rotational symmetry of the 
lattice.    The collinear stripe state also breaks translational symmetry in the direction perpendicular to the stripes.   
But in the presence of a magnetic easy axis, it {\it does not} break spin rotation symmetry.  

This collinear stripe state supports two branches of spin-wave excitations, which are degenerate in the absence of magnetic field.   
Both branches are gapped.  However for parameters relevant to AgNiO$_2$, the minimum of the spin wave dispersion does {\it not} 
occur at the magnetic ordering vector $M$, as would be expected, but rather at points $M'$ related by the broken 
$\mathds{Z}_3$ rotational symmetry\cite{wheeler09}.

\begin{figure}[ht]
\centering
\includegraphics[totalheight=0.115\textheight]{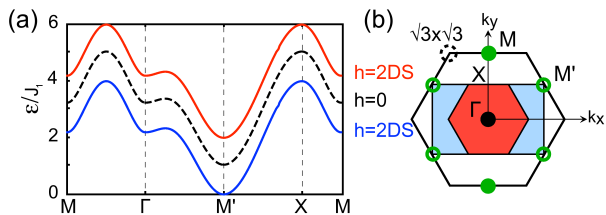}
\caption{\footnotesize{   (Color online) (a) Linear spin-wave dispersion of the collinear stripe phase of Eq.~\ref{eq:H} for $h$=$0$ 
(dashed line) and $h$=$2DS$ (solid lines) in the $k_z$=$0$ plane, for $S=1$, $J_1$=$1$, $J_2$=$0.15$, $J_\perp$=-$0.15$ 
and $D$=$0.5$.
(b) First Brillouin zone for a triangular lattice, showing the ordering vector $M$, and related symmetry points $M^\prime$. 
The magnetic Brillouin zones for the collinear stripe phase and associated supersolid are shown 
by a blue rectangle and a red hexagon, respectively.
}}
\label{lsw}
\end{figure}

Applying a magnetic field parallel to the easy axis lifts the degeneracy of the two spin wave branches, and reduces the gap at $M'$.    
Within linear spin wave theory, neglecting dispersion in the out-of-plane direction and expanding about a stripe state with ordering 
vector $M$=$(0,2\pi/\sqrt{3})$, we find
\begin{eqnarray}
\label{lswgap}
&& \epsilon_\pm(\mathbf{k}) = 4S \left[ \left( J_1 \cos^2 (k_x/2) +J_2\cos^2 (\sqrt{3}k_y/2) + D/2 \right)^2 \right.  \nonumber\\
&& \left. -\left(  \cos(\sqrt{3}k_y/2)\left( J_1 \cos (k_x/2) + J_2 \cos (3k_x/2) \right) \right)^2\right]^\frac{1}{2} \pm  h, \nonumber  
\end{eqnarray}
and the spin gap at $M'$~=$(\pm \pi, \pi/\sqrt 3)$ closes completely at a critical field $h$=$2DS$.   
The resulting dispersion is shown in Fig.~\ref{lsw} (a).  

As in the celebrated example of TlCuCl$_3$, the closing of this spin gap leads to Bose-Einstein condensation of 
spin-wave excitations (magnons)\cite{giamarchi08}.   This Bose-Einstein condensate breaks a $U(1)$ spin-rotation symmetry in the $S^x$-$S^y$ plane, and so has superfluid character.   Since the resulting state inherits the broken $\mathds{Z}_2$ translational and $\mathds{Z}_3$ rotational symmetries 
of the collinear ground state, it is a supersolid.  This quantum phase transition can also be understood at a mean-field level --- instead of undergoing a spin-flop, the ``down'' spins cant, while the ``up'' spins remain aligned with the field.   
The nature of the new magnetic supersolid is illustrated in Fig.~\ref{cartoon}(b).  

These arguments establish the possibility of a supersolid state in AgNiO$_2$, but tell us nothing about its thermodynamic properties.
If the low-field transition in AgNiO$_2$ is into a supersolid, why does the critical field increase with increasing temperature ?
What might the experimental signatures of this new phase be   ?
What other states might occur at higher magnetic field, and how do they evolve with temperature ?

In order to address these questions, we have performed classical Monte Carlo simulations of Eq.~\ref{eq:H}.   
The combination of large magnetic anisotropy and competing interactions means that 
simulations based on a Metropolis algorithm with a simple local update suffer severe freezing.  
To overcome these problems we employed a parallel tempering Monte Carlo scheme\cite{hukushima96}, combined 
with successive over-relaxation sweeps\cite{kanki05}.  
Simulations of 48-128 replicas were performed for rhombohedral clusters of 3$L$$\times$$3L$$\times$$L$=$9L^3$ spins, 
where $L$=4,6,8,10 counts the number of triangular lattice planes.    
Periodic boundary conditions were imposed.    
Typical simulations involved 4$\times$10$^6$ steps, half of which were discarded for thermalization.   
Each step consisted of one local-update sweep of the lattice, followed by 
two over-relaxation sweeps, with replicas at different temperatures exchanged every 10 steps.
We set $|{\bf S}| = S =1$ throughout.  

\begin{figure}[ht]
\centering
\includegraphics[totalheight=0.4\textheight]{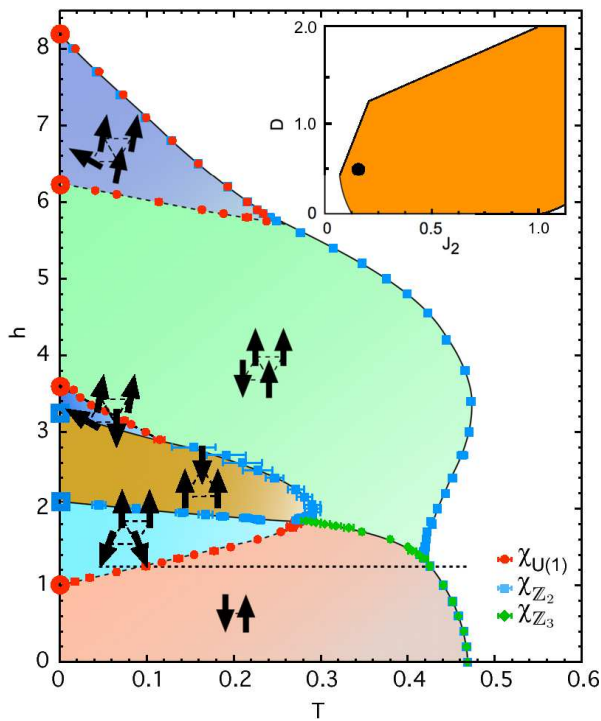}
\caption{\footnotesize{(Color online)  Magnetic phase diagram obtained from classical Monte Carlo simulation of Eq.~\ref{eq:H} 
for a cluster of 24$\times$24$\times$8  spins  with $J_1$=$1$, $J_2$=$0.15$, $J_\perp$=-$0.15$, $D$=$0.5$.  
Temperature $T$ and magnetic field $h$ are measured in units of $J_1$.   
Phase boundaries are determined from peaks in corresponding order parameter susceptibilities.   Phase transitions
are first order, except where shown with a dashed line.  
A dotted black line shows the cut at $h$=$1.25$ used in Fig.~\ref{transitions}(a)--(d).   Inset shows the range of 
parameters for which a supersolid arises as the first instability of the stripe phase in magnetic field, 
as determined by mean field calculations for $J_\perp$=-$0.15$. 
}}
\label{phasediagram}
\end{figure}

\begin{figure}[ht]
\centering
\includegraphics[width=0.37\textheight]{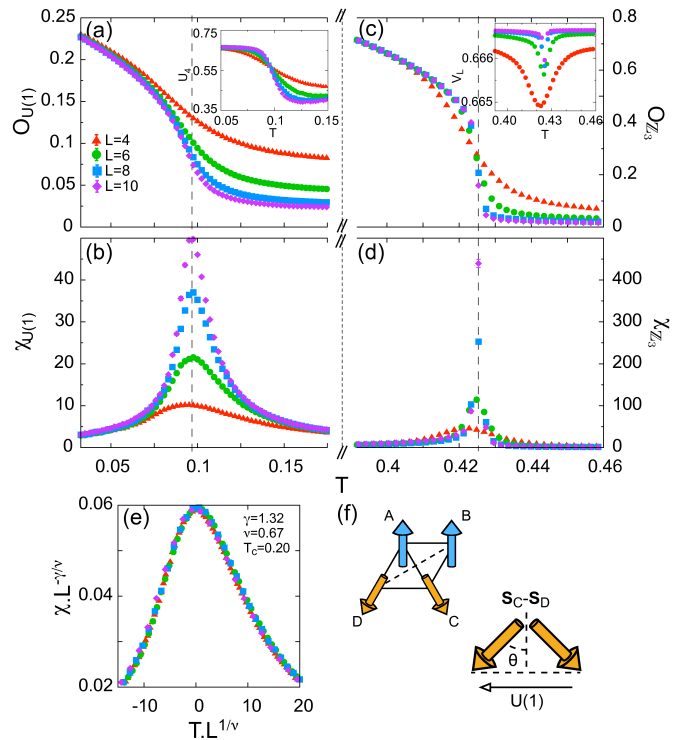}
\caption{\footnotesize{(Color online) 
a) $U(1)$ order parameter showing onset of supersolid phase for $h$=$1.25$, $T$$\approx$$0.1$ (inset : crossing of associated Binder cummulants).   b) related order parameter susceptibility $\chi_{U(1)}$.  
c)~$\mathds{Z}_3$~order parameter showing onset of collinear stripe phase for $h$=$1.25$, $T\approx 0.42$  
(inset :  Binder cumulants for energy, showing a dip indicative of a bimodal distribution).
d) related order parameter susceptibility $\chi_{\mathds{Z}_3}$. 
Results are taken from simulations of clusters of $3L$$ \times$$ 3L$$\times$$L$ spins, with $L$=4, 6, 8, 10, for  
parameters identical to Fig.~\ref{phasediagram}.  
e) finite-size scaling of order parameter susceptibility at transition into supersolid for $h$=$1.5$, $T$$\approx$$ 0.2$.  
f) graphical representation of U(1) order parameter as a vector in the $S^x$-$S^y$ plane.  }}
\label{transitions}
\end{figure}

The results of these simulations are summarised in Fig.~\ref{phasediagram}.   For the parameters used, we find a total of six 
distinct phases as a function of increasing field : 
i) a collinear ``stripe'' ground state with a 2-site unit cell; 
ii) a supersolid phase with a 4-site unit cell, continuously connected with (i); 
iii) a collinear one-third magnetization plateau state with a 3-site $\sqrt{3}\times\sqrt{3}$ unit cell; 
iv) a second supersolid, formed by a 2:1:1 canting of spins within the same 4-site unit cell as (ii);
v) a collinear half-magnetization plateau state, with the same 4-site unit cell as (ii); 
and 
vi) a third supersolid, with the same 4-site unit cell as (ii), formed by a 3:1 canting of spins approaching saturation.  
Phase transitions were identified using peaks in the relevant order-parameter susceptibilities.   
These transitions are generically first order, except between collinear and supersolid phases with the same 
unit cell, where transitions are found to be continuous.   
All of these phases can also be found in mean field theory at $T=0$, and transitions between them are shown 
by open squares, diamonds or circles on the h-axis of Fig.~\ref{phasediagram}. 
 
This phase diagram shows some intriguing similarities with experimental work on AgNiO$_2$\cite{coldea10}.  
In particular, the topology of the low-field phases is correctly reproduced, with the low-field supersolid phase contained entirely within 
the envelope of the collinear stripe phase.   The phase transition between these two phases is continuous, and the critical field increases 
with increasing temperature\footnote{At present, we do not find any evidence for the more delicate Berezinskii-Kosterlitz-Thouless transitions 
found in the two-dimensional easy-axis magnets --- cf. P.-\'E. Melchy and M. E. Zhitomirsky, Phys. Rev. B {\bf 80}, 064411 (2009), 
and references therein.}.  
For this reason we now concentrate on the low-field properties of the model, leaving the rich physics at higher field for discussion elsewhere.  
We note, however, that the one-third magnetization plateau (iii) is well known from studies of 
easy-axis triangular lattice antiferromagnets\cite{miyashita86}, and that states analogous to 
the half-magnetization plateau (v) and high field supersolids (iv) and (vi) 
%(iv), (v) and (vi) 
also occur in models of Cr spinels\cite{penc04}.    

As in some previously studied models\cite{melko05,laflorencie07}, two finite-temperature phase transitions separate the low-field supersolid 
phase from the paramagnet.   The first of these is a first-order transition into the collinear stripe state at a temperature 
$T \approx  0.42$.   
The second is a continuous transition at a critical temperature which varies approximately linearly with magnetic field 
from $T=0$ ($h=1$) to $T\approx  0.3$ ($h\approx 2$).    Both translational and rotational lattice symmetries are broken 
at the upper transition.   To study this it is convenient to introduce a two-component order parameter based on an 
irreducible representation of the $C_3  \cong \mathds{Z}_3$ rotation group, which measures the orientation of the 
``stripes'' in the plane
\begin{eqnarray}
\psi_{2s} &=& %1/(\sqrt{6}N)
\frac{1}{\sqrt{6}N}
\sum_i 2S_i^zS_{i+\delta_1}^z-S_i^zS_{i+\delta_2}^z-S_i^zS_{i-\delta_1-\delta_2}^z, \nonumber\\
\psi_{2a} &=&%-i/(\sqrt{2}N)
-\frac{i}{\sqrt{2}N}
\sum_i S_i^zS_{i+\delta_2}^z-S_i^zS_{i-\delta_1-\delta_2}^z, \nonumber
\end{eqnarray}
where ${\bf \delta}_1 = (1,0)$ and ${\bf \delta}_2 = (1/2,\sqrt{3}/2)$ are the primitive vectors of the triangular lattice.
Figure~\ref{transitions}(c) and (d) show the behaviour of this order parameter, Binder cumulants for energy, and related susceptibility for $h=1.25$, $T \approx 0.42$.
The transition is clearly first order; we have checked explicitly that the $\mathds{Z}_2$ symmetry associated with translations is broken at the 
same temperature.

\begin{figure}[ht]
\centering
\includegraphics[width=0.35\textheight]{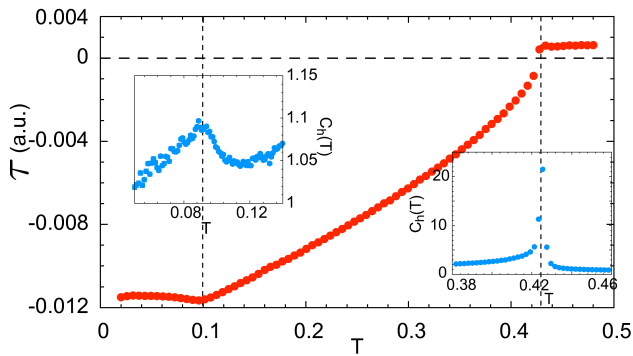}
\caption{\footnotesize{(Color online) 
Magnetic torque $\tau = {\bf m} \times {\bf h}$ as a function of temperature, for a magnetic field of magnitude 
$1.25$ at an angle of $5^o$ to the easy axis (natural units).  
Torque changes sign abruptly at the first order transition from paramagnet to collinear stripe phase for $T \approx 0.42$. 
A change in slope for $T \approx 0.1$ signals the continuous transition from stripe phase to magnetic supersolid.   
Insets show the heat capacity anomalies at each of these transitions.
}}
\label{torque}
\end{figure} 

Spin rotation symmetry in the $S^x$-$S^y$ plane is broken at the lower phase transition into the supersolid state.   
This can be measured by constructing a $U(1)$ order parameter which measures the difference between $S^x$ and $S^y$ components 
of the canted ``down'' spins, as illustrated in Fig.~\ref{transitions}(f).   Fig.~\ref{transitions}(a) and (b) show the behaviour of this order parameter, its Binder cumulants, and related susceptibility for $h=1.25$, $T \approx 0.1$.    The phase transition 
remains continuous at finite temperature, with Binder cumulants for different system size crossing at a single temperature.    
A good collapse of susceptibility data is obtained using susceptibility and correlation length exponents $\gamma = 1.32$ 
and $\nu = 0.67$ for the 3D XY universality class, as shown in Fig.~\ref{transitions}(e).

While the relative extent of each phase and details of critical fields and temperatures are different, AgNiO$_2$ exhibits
a similar double transition on cooling; a transition from paramagnet to a collinear stripe phase at $T$$\approx$20K
accompanied by a sharp feature in specific heat and then, for fields greater than 13.5T, a continuous
or very weakly first-order transition from the collinear stripe phase into an unknown low temperature magnetic state.  
This transition occurs at higher magnetic field for higher temperatures, suggesting that the stripe phase 
has higher entropy than the competing high-field phase, as found in our simulations.   
Is the high field phase in AgNiO$_2$ then a supersolid ?

Direct confirmation of the magnetic order in AgNiO$_2$ for fields greater than 13.5T by elastic neutron scattering is 
feasible, but challenging, since no large single crystals are presently available.    None the less, it should be possible to 
observe the closing of the spin gap on entry to the supersolid phase in inelastic neutron scattering experiments
on powder samples\cite{wheeler09}.    Moreover, both transport and thermodynamic measurements on small single crystals 
clearly resolve magnetic phase transitions in AgNiO$_2$ \cite{coldea10}.   We therefore conclude by examining 
the heat capacity and magnetization (torque) signatures of the stripe and supersolid phases of our model.  

In Fig.~\ref{torque} we present predictions for magnetic torque, ${\bf \tau} = {\bf m} \times {\bf h}$, and heat capacity $C_h$, 
for the same fixed value of magnetic field $h=1.25$ chosen for the study of phase transitions in Fig.~\ref{transitions}.  
Torque changes sign at the first-order phase transition from the paramagnet at $T\approx 0.4$, 
is strongly temperature dependent in the stripe phases for $0.1 \lesssim T \lesssim 0.4$, 
and is broadly temperature-independent in the supersolid phase for $T\lesssim 0.1$\footnote{
These predictions should be contrasted with earlier work on torque at spin-flop transition : 
T. Nagamiya {\it et al.}, Adv. Phys. \textbf{13}, 4 (1955).}.  
For the ratios of parameters used, with $J_1$$=$1.32meV (cf.\cite{wheeler09}), 
this translates into a supersolid transition at a field of 12.5 Tesla, for a temperature of 1.5K.
The heat capacity anomalies at both transitions strongly resemble those observed in AgNiO$_2$\cite{coldea10}.  

In summary, we have studied the magnetic phase diagram of a realistic three-dimensional spin model 
with single-ion anisotropy and competing interactions on a layered triangular lattice, identifying three phases 
which are magnetic supersolids in the sense of Liu and Fisher\cite{liu73}.    We find that these supersolids 
are continuously connected with parent collinear phases through the Bose-Einstein condensation of 
magnons.   However, quantum fluctuation effects may also play an important role at these phase transitions, 
and this remains an interesting problem for future study.
The model studied was motivated by the metallic triangular lattice antiferromagnet \mbox{$2H$-AgNiO$_2$}, and is known 
to describe its magnetic excitations in zero field\cite{wheeler09}.
Since the model does not take 
itinerant charge carriers into account, it cannot pretend to be a complete theory of AgNiO$_2$.   
Nonetheless, it motivates a re-examination of the low field transitions observed in AgNiO$_2$, where 
a magnetic supersolid may already have been observed\cite{coldea10}.  

%%%%%%%%%%%%%%%%%%%%%%%%%%%%%%%%%%%%%%%%%%%

{\it Acknowledgments:} The authors thank Pierre Adroguer, Tony Carrington, Amalia and Radu Coldea, 
Andreas L\"auchli, Yukitoshi Motome and Mike Zhitomirsky for many helpful comments on this work.
Numerical simulations made use of the Advanced Computing Research Centre, University of Bristol.   
This work was supported by FCT fellowship SFRH/BD/27862/2006 and EPSRC Grants EP/C539974/1
and EP/G031460/1.

%%%%%%%%%%%%%%%%%%%%%%%%%%%%%%%%%%%%%%%%%%%%%%%%%%%%%%


\begin{thebibliography}{99}
    
\bibitem{andreev69} 
   A.F. Andreev and I.M. Lifshitz, Sov Phys. JETP \textbf{29}, 1107 (1969); 
   G. Chester, Phys. Rev. A \textbf{2}, 256 (1970); T. Leggett, Phys. Rev. Lett. \textbf{25}, 1543 (1970).
\bibitem{chan04} E. Kim and M.H.W. Chan, Nature \textbf{427}, 225 (2004).
\bibitem{liu73}  K.S. Liu and M.E. Fisher, J. Low Temp. Phys. \textbf{10}, 655 (1973).
   % other papers ?
\bibitem{melko05} 
   S.~Wessel and M. Troyer, Phys. Rev. Lett. \textbf{95}, 127205 (2005).
   D.~Heidarian and K. Damle, Phys. Rev. Lett. \textbf{95}, 127206 (2005); 
   R.G. Melko \textit{et al.}, Phys. Rev. Lett. \textbf{95}, 127207 (2005); 
   M.~Boninsegni and N. Prokof'ev, Phys. Rev. Lett. \textbf{95}, 237204 (2005). 
\bibitem{wawrzynska07} 
   E. Wawrzynska, \textit{et al.}, Phys. Rev. Lett. \textbf{99} 157204 (2007);
   Phys Rev B \textbf{77} 094439 (2008).
\bibitem{coldea10} A. Coldea {\it et al.}, preprint arXiv:0908.4169v1 (2009).
\bibitem{wheeler09} E. M. Wheeler {\it et al.}, Phys Rev B \textbf{79}, 104421 (2009).
\bibitem{giamarchi08} 
   See, e.g., T. Giamarchi {\it et al.}, N. Phys. {\bf 4}, 198 (2008).
\bibitem{laflorencie07}  N. Laflorencie and F. Mila, Phys. Rev. Lett. \textbf{99}, 027202 (2007).
\bibitem{hukushima96} K. Hukushima and K. Nemoto, J. Phys. Soc. Jpn. \textbf{65}, 1604 (1996).
\bibitem{kanki05} K. Kanki {\it et al.}, Eur. Phys. J. B \textbf{44}, 309 (2005).
\bibitem{miyashita86} S. Miyashita, J. Phys. Soc. Jpn. \textbf{55}, 3605 (1986).
\bibitem{penc04} K. Penc, N. Shannon and H. Shiba, Phys. Rev. Lett. {\bf 93}, 197203 (2004).
\bibitem{nagamiya55} T. Nagamiya {\it et al.}, Adv. Phys. \textbf{13}, 4 (1955).

\end{thebibliography}
\end{document}